\documentclass[prl,aps,amsmath,amssymb,twocolumn,groupedaddress ,floatfix,hidelinks]{revtex4-2}
\usepackage{graphicx,
	amsmath,
	braket,
	units,
	ragged2e,
	xcolor,
	xspace,
	nicefrac,
	lipsum,
	nameref,
	textcomp,
	urwchancal,
	upgreek,
	isotope,
	hyperref,
	cleveref,
	placeins, 
	scrextend, 
	url}
\usepackage{amstext}
\usepackage{amsfonts}
\usepackage{amsxtra}
\usepackage{bm}
\usepackage{grffile}
\usepackage{soul}
\usepackage{epstopdf}
\usepackage{marvosym} 
\usepackage{tikz}
\usepackage{flushend}
\usepackage{scalerel}

\usetikzlibrary{svg.path}

\hypersetup{
	colorlinks=true,
	allcolors = blue
}

\DeclareMathAlphabet{\mathcal}{OMS}{cmsy}{m}{n}

\newcommand{\rr}{\mathbf{r}}

\newcommand{\edd}{\varepsilon_\text{dd}}

{%
	\setlength{\fboxsep}{0pt}%
	\setlength{\fboxrule}{1pt}%
}%

\definecolor{orcidlogocol}{HTML}{A6CE39}
\tikzset{
	orcidlogo/.pic={
		\fill[orcidlogocol] svg{M256,128c0,70.7-57.3,128-128,128C57.3,256,0,198.7,0,128C0,57.3,57.3,0,128,0C198.7,0,256,57.3,256,128z};
		\fill[white] svg{M86.3,186.2H70.9V79.1h15.4v48.4V186.2z}
		svg{M108.9,79.1h41.6c39.6,0,57,28.3,57,53.6c0,27.5-21.5,53.6-56.8,53.6h-41.8V79.1z M124.3,172.4h24.5c34.9,0,42.9-26.5,42.9-39.7c0-21.5-13.7-39.7-43.7-39.7h-23.7V172.4z}
		svg{M88.7,56.8c0,5.5-4.5,10.1-10.1,10.1c-5.6,0-10.1-4.6-10.1-10.1c0-5.6,4.5-10.1,10.1-10.1C84.2,46.7,88.7,51.3,88.7,56.8z};
	}
}

\newcommand\orcidicon[1]{\href{https://orcid.org/#1}{\mbox{\scalerel*{
				\begin{tikzpicture}[yscale=-1,transform shape]
					\pic{orcidlogo};
				\end{tikzpicture}
			}{|}}}}

\newcommand{\vv}{\mathbf{v}}

\begin{document}
	\title{Analytic Phase Solution and Point Vortex Model for Dipolar Quantum Vortices}
	
	\author{Ryan Doran\,\orcidicon{0000-0002-9467-1264}}
	\email{ryan.doran@newcastle.ac.uk}
	\affiliation{Joint Quantum Centre (JQC) Durham--Newcastle, School of Mathematics, Statistics and Physics, Newcastle University, Newcastle upon Tyne, NE1 7RU, UK}
	
	\author{Thomas Bland\,\orcidicon{0000-0001-9852-0183}} \email{thomas.bland@fysik.lu.se}
	\affiliation{Mathematical Physics, LTH, Lund University, Post Office Box 118, S-22100 Lund, Sweden}
	
	\date{\today}
	
	\begin{abstract}
		We derive an analytic expression for the phase of a quantum vortex in a dipolar Bose–Einstein condensate, capturing anisotropic effects from long-range dipole–dipole interactions. This solution provides a foundation for a dipolar point vortex model (DPVM), incorporating both phase-driven flow and dipolar forces. The DPVM reproduces key features of vortex pair dynamics seen in full simulations, including anisotropic velocities, deformed orbits, and directional motion, offering a minimal and accurate model for dipolar vortex dynamics. Our results open the door to analytic studies of vortices in dipolar quantum matter and establish a new platform for exploring vortex dynamics and turbulence in these systems.
	\end{abstract}
	\maketitle

	Circulating flow structures, or vortices, are central to the dynamics of both classical and quantum fluids. In systems governed by quantum coherence, such as superfluids and superconductors, vortices arise not as turbulent eddies but as topological singularities in the phase of a complex order parameter. Around each vortex core, the phase winds by integer multiples of $2\pi$, enforcing quantized circulation and constraining the velocity field to be irrotational except at isolated points~\cite{Feynman1955cia}. These features make quantum vortices not only signatures of superfluidity and superconductivity, but also essential building blocks for understanding quantum turbulence \cite{henn2009emergence,tsatsos2016quantum,madeira2020quantum}, lattice formation \cite{abrikosov2004nobel,aboshaeer2001observation,Fetter2009}, and long-range coherence in interacting many-body systems \cite{Zwierlein2005vas,Casotti2024oov,poli2025synchronization}.
	
	Quantum vortices have a fixed core size and quantized circulation, meaning that the can be effectively described as interacting point particles. The point vortex model (PVM) offers a numerically efficient framework for capturing the dynamics of well-separated vortices in an ideal two-dimensional fluid \cite{Helmholtz1858}. While originally developed for classical incompressible flows, the model proves remarkably accurate for superfluid systems. The PVM has successfully described vortex dynamics in superfluid He \cite{Sachkou2019coherent}, superfluid Fermi gases \cite{Kwon2021sound,hernandez2024connecting}, and atomic Bose–Einstein condensates (BECs) \cite{Gauthier2019,Reeves2022}. This has led to a number of profound results in superfluid dynamics, from vortex-sound interaction \cite{Kwon2021sound} and vortex instabilities \cite{hernandez2024connecting} to the enstrophy cascade \cite{Reeves2017_enstrophy} and the prediction \cite{Billam2014_OK} and experimental realization of large-scale flows \cite{Reeves2022} and Onsager vortex clustering \cite{Gauthier2019}.
	
	The key ingredient in the point vortex model is the phase profile of a single vortex. For quantum gases that interact only through short-range contact-interactions, this phase has the simple analytic form $S(x,y)=\arctan(y/x)$ \cite{Fetter1966a}. However, in certain classes of quantum matter with anisotropic and long-range interactions---such as dipolar quantum fluids---the interaction itself breaks radial symmetry, and no closed-form expression for the vortex phase is currently known~\cite{Martin2017}. Vortices in dipolar BECs were observed only recently \cite{Klaus2022oov}, made possible by the development of magnetostirring techniques that exploit the anisotropic dipolar interaction \cite{Prasad2019vlf,Bland2023vid}. This experiment revealed striking vortex stripe lattices, a hallmark of anisotropic vortex–vortex interactions \cite{Zhang2005vli,Ticknor2011asi,Prasad2019vlf}. Several other theoretical predictions remain unconfirmed, including elliptically distorted vortex cores, roton-induced density oscillations around the core \cite{Yi2006vsi,Ticknor2011asi,JonaLasinio2013rci,JonaLasinio2013tof,Mulkerin2013aal,mulkerin2014vortices,Bland2023vid,Prasad2024vortex,villois2025vortex}, and anisotropic vortex pair dynamics such as direction-dependent motion and suppressed annihilation \cite{Ticknor2011asi,Mulkerin2013aal,gautam2014dynamics,Prasad2024vortex}.
	
	In this Letter, we take a first-principles approach to derive the phase profile of a quantum vortex in a dipolar condensate. Our analytic solution matches the numerically obtained phase with high accuracy, and allows us to approximate the density profile of the vortex core. We use this result to formulate the dipolar point vortex model (DPVM), a minimal model that incorporates anisotropic interactions through an analytic phase field and a phenomenological inclusion of the long-range dipole-dipole interaction between vortices \footnote{We note here that throughout this work ``dipole'' refers to the long-range interaction, not to be confused with vortex dipoles, which we will refer to as vortex-antivortex pairs.}. The DPVM is rigorously benchmarked against mean-field simulations of vortex pair dynamics, reproducing key features such as elliptic vortex–vortex orbits and directionally dependent motion of vortex–antivortex pairs \cite{Ticknor2011asi,Mulkerin2013aal,gautam2014dynamics,Prasad2024vortex}.
	
	{\it Deriving the dipolar vortex phase}---In an infinite, homogeneous two-dimensional system, the phase $S(\rho,\vartheta)$ of a vortex with density $n(\rho,\vartheta)$ in an irrotational BEC must satisfy the continuity equation~\cite{Fetter1966a,Berloff2004_pade,Bradley2012}
	\begin{align}
		\frac{\partial n}{\partial t} + \nabla\cdot\left(n \, \nabla S\right)=0\,.
		\label{eqn:continuity_equation}
	\end{align}
	Assuming the phase is independent of radius, $\rho = \sqrt{x^2 + y^2}$, and the density is independent of the azimuthal angle, $\vartheta = \arctan(y/x)$, Eq.~\eqref{eqn:continuity_equation} reduces trivially to $\partial^2 S(\vartheta) / \partial \vartheta^2 = 0$, which has solution $S(\vartheta) = q\vartheta$, where $q \in \mathbb{Z}$ is the topological charge of the vortex~\cite{Fetter1966a}. The second constant of integration is a phase shift that can be set to zero. This procedure is exact for any superfluid satisfying these assumptions, up to the singularity at $\rho = 0$.
	
	\begin{figure}
		\centering
		\includegraphics[width=1\linewidth]{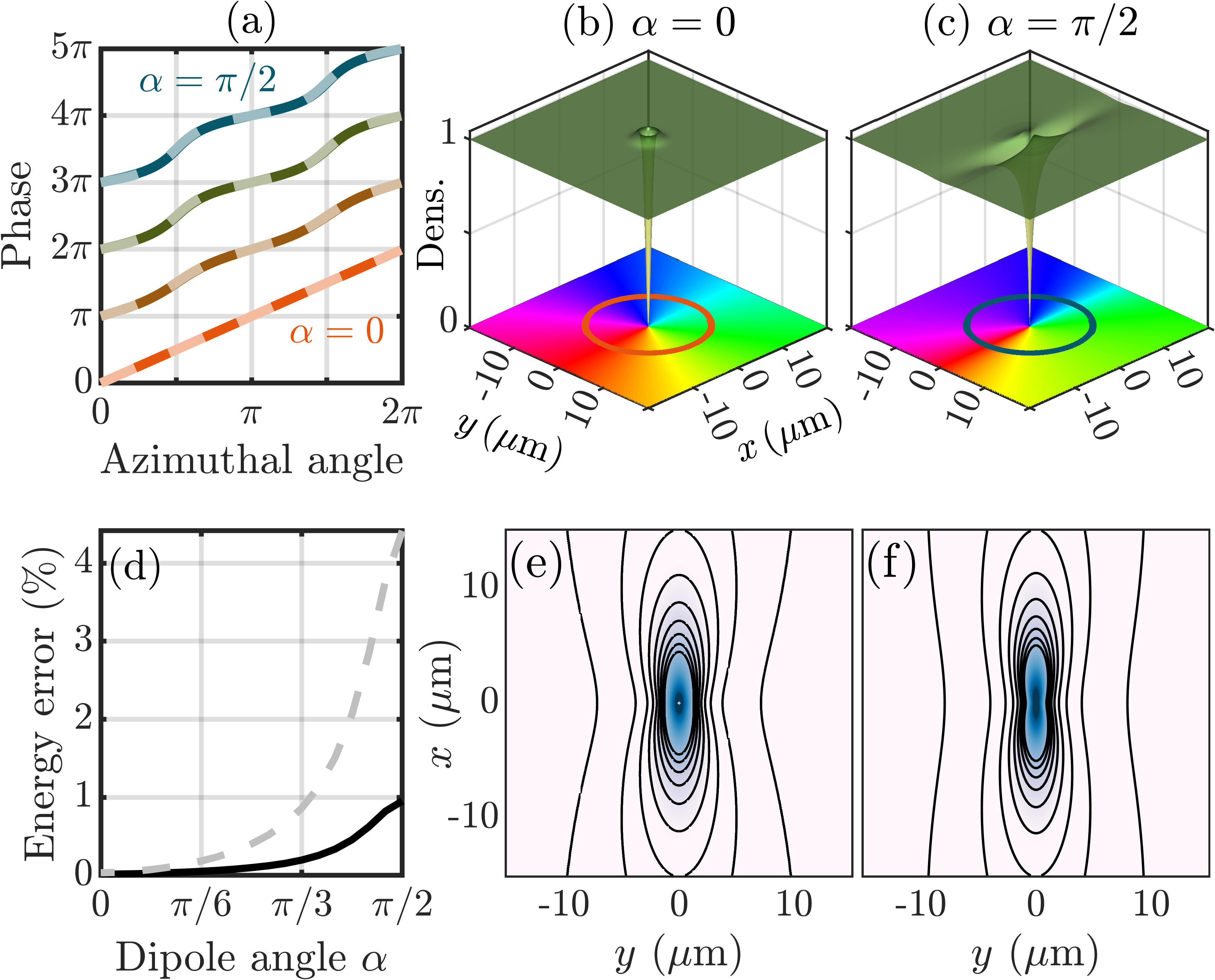}
		\caption{Analytic phase of an elliptic vortex. (a) Solid lines show the vortex phase obtained from the dipolar GPE with interaction strength $\edd=0.9$ at fixed radius $R = 10\,$\textmu{m}; dashed lines indicate the corresponding solution to Eq.~\eqref{eqn:elliptic_phase}, using only the fitted ellipticity $\lambda$ extracted from the vortex density. The polarization angle $\alpha$ increases in steps of $\pi/6$, and each curve is vertically offset by $\pi$ for clarity. (b,c) Example solutions for the normalized vortex density $n(x,y)$ and phase at $\alpha = 0$ and $\alpha = \pi/2$, respectively. The colored circle marks the location from which the phase is extracted in (a). (d) Relative energy error: GPE vs. vortex Ansatz (Eq.~\eqref{eqn:density_Ansatz}, solid), and GPE vs. isotropic Ansatz ($\lambda = 1$, dashed). Speed $|\nabla S|$ with $\alpha=\pi/2$ from the (e) GPE and (f) analytic solution, with contours at steps of 20\textmu{m}/s.}
		\label{fig:one_vortex}
	\end{figure}
	
	In dipolar systems, however, vortices are predicted to develop elliptic cores~\cite{Yi2006vsi,Ticknor2011asi,JonaLasinio2013rci,JonaLasinio2013tof,Mulkerin2013aal,mulkerin2014vortices,Bland2023vid,Prasad2024vortex}, accompanied by lobes of increased and depleted density extending away from the center, invalidating the assumptions above. This anisotropic structure arises when the polarization axis is tilted by an angle $\alpha$ relative to the vortex line (normal to the $xy$-plane), resulting in magnetostriction~\cite{Ekreem2007aoo}--a stretching of the vortex core along the direction of polarization. Exemplar dipolar vortex density and phase profiles are shown in Fig.~\ref{fig:one_vortex}, obtained as stationary solutions of the dipolar Gross-Pitaevskii equation, which will be discussed in detail later. When the polarization axis is normal to the plane ($\alpha=0$), the dipoles form an azimuthally symmetric density ring around the core, and the phase is unchanged from the non-dipolar result, Fig.~\ref{fig:one_vortex}(b). However, tilting the dipoles into the plane breaks the angular symmetry of the density. In the strongly dipolar case with maximum tilt angle $\alpha=\pi/2$, as shown in Fig.~\ref{fig:one_vortex}(c), the core width broadens along the polarization direction, while regions of high density are maintained perpendicular to it. Strikingly, the phase profile responds to this deformation by concentrating phase gradients--hence superfluid velocity--along the polarization axis, with diminished gradients perpendicular to it. This is seen clearly by examining the non-linear phase at a fixed radius away from the core, as shown in Fig.~\ref{fig:one_vortex}(a).
	
	\begin{figure*}[!ht]
		\centering
		\includegraphics[width=\linewidth]{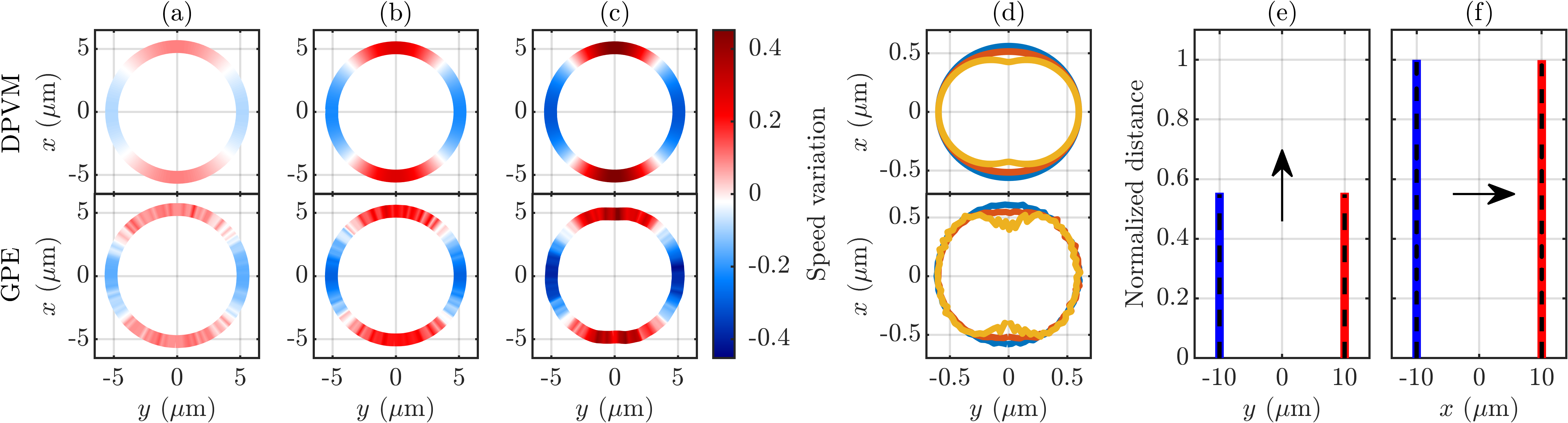}
		\caption{Comparison of vortex pair dynamics. (a-c) Vortex-vortex orbits with initial condition $x_j=0$ and $y_1=-y_2=5$\textmu{m} in the DPVM (top row) and GPE (bottom row) for $\edd=0.9$ and (a) $(\alpha,\lambda)=(\pi/6,1.08)$ (b) $(\pi/3,1.15)$ and (c) $(\pi/2,1.3)$. Color represents relative deviation from the local speed to the average speed $v$, i.e.~$(|\textbf{v}(x,y)| -v)/v $. (d) Same as (a-c) with $y_1=-y_2=0.6$\textmu{m}, with fixed $\alpha=\pi/2$ and $(\edd,\lambda)=(0.3,1.03)$ (0.6,1.08) (0.9,1.15). (e-f) Traveling vortex-antivortex pairs with the vector between them (e) perpendicular and (f) parallel to the magnetic field and $(\edd,\alpha,\lambda)=(0.9,\pi/2,1.3)$. The DPVM vortex (blue, $q=1$) and antivortex (red, $q=-1$) are shown with the corresponding GPE solutions (black dashed lines) overlaid, normalized to the same maximum traveled distance in (f). We fix $\xi_v = 20.3a_\text{dd}$ throughout.}
		\label{fig:two_vortices}
	\end{figure*}
	
	To find the dipolar vortex phase, it is natural to search for solutions to Eq.~\eqref{eqn:continuity_equation} in a modified coordinate system. By making the scale transformation $y\to\lambda^{-1} y$ we can define a shear-polar coordinate system ($\rho',\vartheta'$) where $x=\rho'\cos\vartheta'$ and $y=\lambda^{-1}\rho'\sin\vartheta'$. Circles in $(\rho',\vartheta')$ are ellipses in ($\rho,\vartheta$), hence $\lambda>1$ parametrizes the ellipticity of the vortex core. This transformation precisely recovers the assumption $S(\rho',\vartheta')\equiv S(\vartheta')$. We make a further approximation that in this coordinate system the vortex density is independent of $\vartheta'$, which we expect to be true away from the roton-induced density oscillations. Carefully defining the $\nabla'$ operator in this coordinate system (see End Matter) the stationary continuity equation is
	\begin{align}
		(\nabla')^2 S = \frac{\partial}{\partial \vartheta'} \left[ \sqrt{\frac{\tan^2\vartheta'+\lambda^2}{1+\lambda^2\tan^2\vartheta'}} \frac{\partial}{\partial \vartheta'} \right]S(\vartheta') = 0\,,
	\end{align}
	which is valid everywhere except at the origin. Remarkably, this equation is solvable and leads us to a solution for the phase. After integration and transformation back to Cartesian coordinates, we obtain 
	\begin{align}    \label{eqn:elliptic_phase}
		S(x,y) &= q \Lambda \Bigg\{ \left(\lambda^2-1\right) F\left[\arctan\left(\frac{y}{x}\right)\, \Bigg| 1-\lambda^4\right] \nonumber  \\ & \ + \lambda^2 \Pi\left[1-\lambda^2\,;\arctan\left(\frac{y}{x}\right)\,\Bigg|1-\lambda^4\right] \Bigg\} \,,
	\end{align}
	where $F(z|m)$ and $\Pi(n; z|m)$ are the incomplete elliptic integrals of the first and third kinds, respectively~\cite{Abramowitz1988Handbook}. The integration constant is $\Lambda = \pi \left[2\lambda^{-2} \Pi\left(1 - \lambda^{-2} \big| 1 - \lambda^{-4} \right) - 4 K\left(1 - \lambda^{-4}\right) \right]^{-1}$, where $K(m)\equiv F(\pi/2|m)$ and $\Pi(n|m)$ denote the corresponding complete first and third elliptic integrals. This choice ensures periodicity of the solution, $S|_{\vartheta=\pi} = S|_{\vartheta=-\pi}$. We can use this phase profile to approximate the form of the vortex density. We use our shear-polar coordinate system $\left(\rho^\prime,\vartheta^\prime\right)$ to modify the core profile in Refs.~\cite{Fetter1966a,Bradley2012}, which leads to an Ansatz profile for the vortex 
		\begin{equation}
			\psi_v(x,y) = \sqrt{n_0} \frac{\sqrt{x^2+\lambda^2y^2}}{\sqrt{x^2 + \lambda^2y^2 + a^2 }}e^{iS(x,y)}\,,
			\label{eqn:density_Ansatz}
		\end{equation}
		where $n_0$ is the background density of the fluid, and $a$ is a constant setting the core size.
	
	To verify Eq.~\eqref{eqn:elliptic_phase} and Eq.~\eqref{eqn:density_Ansatz}, we introduce the governing equations for ultracold, weakly interacting, dilute dipolar quantum fluids. In a homogeneous magnetic field, interactions are described by the two-body pseudo-potential
		$U(\rr) = g\delta(\rr) + \frac{3g_\mathrm{dd}}{4\pi}\frac{1 - 3\cos^2\theta}{|\rr|^3}$
		for particles of mass $m$. The contact interaction $g = 4\pi\hbar^2 a_\mathrm{s}/m$ depends on the $s$‑wave scattering length $a_\mathrm{s}$, tunable via Feshbach resonances~\cite{Chin2010fri}, while the dipolar term $g_\mathrm{dd} = 4\pi\hbar^2 a_\mathrm{dd}/m$ is set by the dipolar length $a_\mathrm{dd}$. We define $\varepsilon_\mathrm{dd} = g_\mathrm{dd}/g$, noting instability for $\varepsilon_\mathrm{dd} \gtrsim 1$~\cite{Lahaye2009tpo}. The angle $\theta$ is between $\rr$ and the polarization axis.
	
	In two dimensions, the full 3D system is well approximated by the quasi-2D variational dipolar Gross--Pitaevskii equation (GPE)~\cite{Baillie2015agt}, which has been applied across a range of dipolar systems~\cite{Baillie2015agt,Ripley2023tds,yapa2025anomalous}. In this model, the dynamics are governed by the dimensionally reduced wave function $\psi(\boldsymbol{\rho}, t)$, defined on the infinite plane $\boldsymbol{\rho} = (x, y)$, and the dipolar GPE takes the form
	\begin{equation}
		i\hbar\frac{\partial\psi}{\partial t} = \left[-\frac{\hbar^2\nabla^2}{2m} + \int\text{d}^2\boldsymbol{\rho}'\, U_{\alpha,\sigma}(\boldsymbol{\rho}-\boldsymbol{\rho}')|\psi(\boldsymbol{\rho}')|^2\right]\psi\,,
		\label{eqn:2DGPE}
	\end{equation}
	where the modified interaction potential $U_{\alpha,\sigma}$ depends on the polarization angle $\alpha$--with $\alpha = \pi/2$ corresponding to polarization along the $x$ axis--and on the variational width $\sigma$ in the tightly confined $z$ direction.  The areal number density $n_\text{2D} = 500\,$\textmu{m}$^{-2}$ and the transverse harmonic confinement $\omega_z = 2\pi \times 167\,\mathrm{Hz}$ are fixed. This model and our numerical methods are described in the End Matter.

	Figure~\ref{fig:one_vortex} shows the results of the phase and total energy comparison. Solving Eq.~\eqref{eqn:2DGPE} in imaginary time with an initial phase $S =\arctan(y/x)$, we recover the expected elliptic vortex cores for $\alpha \ne 0$ along with corresponding modifications to the phase profile. We extract the ellipticity directly from the density by measuring the square root of the ratio of the full width at half maximum (FWHM) along the $x$ and $y$ directions, obtaining $\lambda_\text{FWHM} = [1,\,1.27,\,1.54,\,1.76]$ for $\alpha = \pi[0,\,1/6,\,1/3,\,1/2]$. Using these values of $\lambda$ as the sole input to Eq.~\eqref{eqn:elliptic_phase}, we compute the dashed lines shown in Fig.~\ref{fig:one_vortex}(a), which show exceptional agreement with the numerically obtained phase. The largest deviation occurs for $\alpha = \pi/2$, where the density is no longer uniform at $R = 10\,$\textmu{m}, and the relative phase error reaches 2\%. We have chosen the FWHM-derived value of $\lambda$ to highlight our model as an experimentally accessible method for inferring the phase profile. However, by instead minimizing the difference between the GPE phase and Eq.~\eqref{eqn:elliptic_phase}, we obtain optimal values that reduce the error to below 1\% for all $\alpha$, namely $\lambda_\text{optim} = [1,\,1.28,\,1.64,\,1.98]$ for the parameters considered. This is illustrated in Fig.~\ref{fig:one_vortex}(d), where using the same $\lambda$ in both the density and phase Ansatz yields a significantly lower relative error in the total vortex energy compared to the standard isotropic Ansatz. Note that the error can be further reduced by considering a smaller value of $\varepsilon_\text{dd}$. Finally, we compare the the velocity fields generated by both methods in Figs.~\ref{fig:one_vortex}(e) and (f). Close to the core, Eq.~\eqref{eqn:elliptic_phase} overestimates the anisotropy, which is also where the density modulations deviate most from a simple ellipse. In the End Matter, we discuss the implications of taking $\lambda(\rho)$.

	{\it The Point Vortex Model---}Having an analytic form that closely matches the phase of an elliptic vortex unlocks the ability to derive a point vortex model for this system. Typically the point vortex model can be derived from the phase of the system \cite{Tornkvist1997}, however as this system is subject to anisotropic long range interactions, it is also necessary to consider the vortex cores, which, as an absence of dipolar fluid, act as dipoles themselves. The velocity contributions to the point vortex model can be decomposed as $\dot{\rr}_j = \vv_j^{(\mathrm{P})} + \vv_j^{(\mathrm{DD})} $,
	where $\dot{\rr}_j$ is the velocity of the $j$-th vortex. The typical point vortex velocity term is proportional to the gradient of the phase,  $\vv_j^{(\mathrm{P})} = \frac{\hbar}{m}\nabla S \big|_{\rr_j}$, while the contribution from the dipole-dipole potential is $\vv_j^{(\mathrm{DD)}} = q_j \boldsymbol{\hat{z}}\times\nabla V_\mathrm{DD} \big|_{\rr_j}$, which is the perpendicular Peach-Koehler term arising from a force acting on the vortex \cite{Peach1950force}. We take the  dipolar potential to be 
	\begin{equation}
		V_\mathrm{DD}(\boldsymbol{\rho}) = \frac{\hbar^2}{m}\frac{\xi_v\edd}{\rho^3}\left(1-3\sin^2\alpha \frac{x^2}{\rho^2}\right)\,,
		\label{eqn:3D_DDI}
	\end{equation}
	where the core size $\xi_v$ accounts for the atoms displaced from the vortex core~\footnote{We have verified that taking the 2D pseudopotential from Ref.~\cite{cai2010mean} does not affect our results.}. From the derivatives of Eqs.~\eqref{eqn:elliptic_phase} and~\eqref{eqn:3D_DDI} we can write the velocity contributions to the dipolar point vortex model as
	\begin{align}
		\label{eqn:vP}
		\vv_j^{(\mathrm{P})} &=  \frac{\hbar\Lambda}{m} \sum_{\substack{k=1 \\ k\neq j}}^{N_v} \frac{q_k}{r_{jk}} \frac{\lambda^4y_{jk}^2 + x_{jk}^2(2\lambda^2-1)}{(x_{jk}^2+\lambda^2y_{jk}^2)\sqrt{x_{jk}^2+\lambda^4y_{jk}^2}} \begin{pmatrix} -y_{jk} \\ x_{jk} \end{pmatrix}\,, \\ 
		\label{eqn:vDD}
		\vv_j^{(\mathrm{DD})} &= \frac{3\hbar^2\xi_v\edd}{m} \sum_{\substack{k=1 \\ k\neq j}}^{N_v} \frac{q_k}{r_{jk}^7} \Bigg\{ \Bigg[5x_{jk}^2\sin^2\alpha - r_{jk}^2\Bigg] \begin{pmatrix} -y_{jk} \\ x_{jk} \end{pmatrix} \nonumber \\ &\qquad\qquad\qquad\qquad\quad- 2\sin^2\alpha\, r_{jk}^2 \begin{pmatrix} 0 \\ x_{jk}  \end{pmatrix}  \Bigg\}\,,
	\end{align}
	where $x_{jk}=x_j-x_k$, $y_{jk}=y_j-y_k$, and $r_{jk}^2=x_{jk}^2+y_{jk}^2$. These expressions clearly reduce to the standard PVM in the non-dipolar limit, $\lambda=1$ and $\edd=0$. 
	
	{\it Vortex-pair dynamics}---With the model in hand, we investigate whether it can capture vortex pair phenomena in dipolar gases. Previous theoretical studies have predicted rich dynamics in such systems, including oval and dumbbell-shaped orbits for vortex–vortex pairs, as well as directionally dependent motion for vortex–antivortex pairs \cite{Ticknor2011asi,Mulkerin2013aal,gautam2014dynamics,Prasad2024vortex,Martin2017}.
	
	In Fig.~\ref{fig:two_vortices}, we present results for both models. We first focus on well-separated vortices separated by 10\textmu{m} that follow near-circular trajectories even in the presence of dipolar interactions. Despite the apparent similarity to the non-dipolar case, the vortex motion is not uniform: the speed peaks when the pair is aligned head-to-tail with the magnetic field direction, and slows down when oriented perpendicular to it. To our knowledge, this anisotropic speed variation for well-separated dipolar vortices has not been discussed in detail beyond its connection to the anisotropic speed of sound~\cite{Ticknor2011asi,Martin2017}, and represents a promising observable for future experimental verification.
	
	To quantify this effect, we plot the instantaneous speed of one vortex along its orbit for increasing magnetic field tilt angles [Fig.~\ref{fig:two_vortices}(a–c)]. The input value of $\lambda$ is obtained by matching the total velocity field from the GPE ground state to $S(\boldsymbol{\rho}-\boldsymbol{\rho}_1)+S(\boldsymbol{\rho}-\boldsymbol{\rho}_2)$ with $S$ given by Eq.~\eqref{eqn:elliptic_phase}, see the End Matter for further discussion. The DPVM captures the angular variation in speed with striking accuracy, exhibiting stronger modulation as the interaction anisotropy increases. Deviations from the GPE results are minimal and primarily attributed to vortex–sound interactions, which introduce visible fluctuations in the numerical simulations. We note that the velocity anisotropy is dominated by the phase contribution, Eq.~\eqref{eqn:vP}.

	Dipolar vortices also exhibit qualitatively distinct orbital shapes. To explore this, we initialize the vortices in close proximity at positions $(0,\pm0.6)$\,\textmu{m}. Inspired by Fig.~3(a) of Ref.~\cite{Prasad2024vortex}, we fix the dipole tilt angle and instead vary the scattering length, thereby tuning the relative dipolar interaction strength $\edd$, as shown in Fig.~\ref{fig:two_vortices}(d). The DPVM again closely reproduces the GPE trajectories, with noticeable deviations appearing only at $\edd = 0.9$, where overlapping density ripples around each vortex push the orbit back toward a circular shape away from head-to-tail configuration. The level of agreement at this short-range scale is especially noteworthy: despite the highly inhomogeneous density in this regime, the essential physics is still accurately captured by the model. In contrast to the velocity modulation discussed above, this orbital deformation is driven by the dipolar interaction term, Eq.~\eqref{eqn:vDD}, highlighting the necessity of its inclusion.
	
	Finally, vortex–antivortex pairs are known to travel perpendicular to their separation vector, and in dipolar condensates this propagation speed additionally depends on the angle between the pair and the magnetic field axis~\cite{Martin2017}. In Figs.~\ref{fig:two_vortices}(e–f), we show that the DPVM captures this anisotropic propagation with high accuracy, matching the net traveled distance to within 1\% of the GPE result.
	
	\begin{figure}
		\centering
		\includegraphics{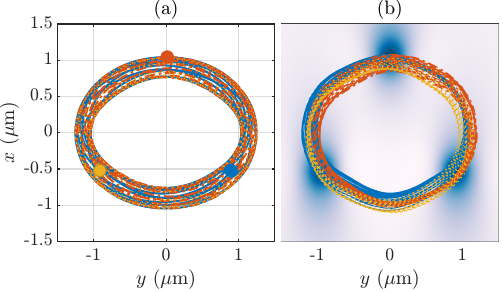}
		\caption{Elliptic orbits of three vortices. Vortices are initially placed at the vertices of an equilateral triangle with circumradius 1 and evolved using (a) the DPVM and (b) the GPE. Vortex trajectories are color- and line-styled as follows: blue, solid; red, dashed; yellow, dotted. Initial vortex positions are marked in (a), while the background in (b) shows the initial condensate density. Parameters: $\alpha=\pi/2$, $\edd=0.6,~\lambda=1.16$, others match Fig.~\ref{fig:one_vortex}.}
		\label{fig:three_vortices}
	\end{figure}
	
	{\it Multi-vortex anisotropic orbits}---An ideal application of the DPVM is the prediction of entirely new classes of stable orbits for $N$-vortex systems. In non-dipolar superfluids, such dynamics have intrigued physicists since Helmholtz's seminal work on vortex motion~\cite{Helmholtz1858,aref2007point}. The canonical example is that of three equally spaced vortices, which form a rigidly rotating triangle--an elegant solution derived analytically by Kirchhoff over two centuries ago~\cite{kirchhoff1883}. In dipolar systems, however, this symmetry is broken by anisotropic interactions. 
	
	Figure~\ref{fig:three_vortices} illustrates a strikingly altered three-vortex orbit. Each vortex now traces a bounded, quasi-periodic path, confined to a torus in phase space. Over time, these paths densely fill out a smooth oval envelope, shown in (a), with no exact repetition. Remarkably, despite the finite core size and density ripples in the full GPE dynamics, the DPVM reproduces the essential features of this motion with high fidelity, as shown in (b). Deviations only arise in regions of strong core–core interactions, where vortex-induced density ripples begin to interfere. 
	
	{\it Conclusions}---We have derived an analytic expression for the phase profile of a single vortex in a dipolar Bose–Einstein condensate by solving the continuity equation in a sheared polar coordinate system. This solution closely matches the phase obtained from mean-field simulations, even in the presence of anisotropy and roton-induced modulations. Using it, we formulated the dipolar point vortex model (DPVM), which incorporates both anisotropic phase gradients and long-range dipole–dipole interactions. The DPVM accurately captures vortex–vortex and vortex–antivortex dynamics seen in full Gross–Pitaevskii simulations, including elliptic orbits and direction-dependent motion. This reduced order model allows us to disentangle the roles played by anisotropic phase gradients and dipole–dipole interactions in governing vortex dynamics.
	
	Our method is general and interaction-agnostic, and may apply directly to systems featuring elliptic vortex/fluxon cores, including Rashba superconductors~\cite{Fuchs2022avs}, coupled light-atom systems~\cite{Leonard2017sfi,Li2017asp,Goldman2014lig}, and molecular superfluids~\cite{Bigagli2024oob}. The framework also extends to any inhomogeneous quantum fluid, provided a coordinate transformation renders the continuity equation solvable. Our model paves the way for efficiently simulating large vortex ensembles—beyond current computational limits~\cite{bland2018quantum,sabari2024vortex,rasch2025anomalous}—where the resulting turbulence is anisotropic, leading to modifications in the vortex decay rate and energy cascade, and enabling the exploration of new stable $N$-vortex configurations in dipolar condensates.
	
	{\it Acknowledgments---}We acknowledge stimulating discussions with Nick Parker during the inception of this project. RD is grateful to Stephanie Reimann for hospitality at Lund University. RD is supported by the UK Engineering and Physical Sciences Research Council, Grant No. EP/X028518/1; TB is supported by the Knut and Alice Wallenberg Foundation (GrantNo. KAW 2018.0217) and the Swedish Research Council (Grant No. 2022-03654VR).

	\newpage
	\onecolumngrid
	\begin{center}{\bf \large End Matter}\end{center}
	\twocolumngrid
	
	{\it Appendix A}---To derive the phase profile of the elliptical vortex, we use the Madelung transform to write the wavefunction as $\psi(\boldsymbol{\rho},t)=\sqrt{n(\boldsymbol{\rho},t)}\exp\left[iS(\boldsymbol{\rho},t)\right]$, where $n$ is the number density of the fluid, and the velocity of the fluid is given by $\vv=(\hbar/m)\nabla S$. To conserve particle number, the density and velocity must obey the continuity equation, Eq.~\eqref{eqn:continuity_equation}. We introduce a shear polar coordinate system that is defined by $x=\rho'\cos\vartheta'$, $y=\lambda^{-1}\rho'\sin\vartheta'$ [therefore $\rho^{\prime 2}=x^2+\lambda^2y^2$ and $\vartheta'=\arctan(\lambda y/x)$], where $\lambda$ parametrizes the ellipticity of the vortex core. We make the approximation that the density profile of the vortex is independent of $\vartheta$ and the phase profile is independent of $\rho$, and look for stationary solutions to the continuity equation; this is akin to saying that we are looking for circular orbits in our new coordinates. We note that in the limit $\lambda=1$ we recover standard polar coordinates, where a density profile with the form $n=f(\sqrt{x^2+y^2})$ and the phase profile $S=\arctan(y/x)$ satisfies the continuity equation. 
	
	In order to find solutions to the continuity equation in this coordinate system, it is necessary to compute the scale factors
	\begin{align}
		h_{\rho'} &= \left| \left(\frac{\partial x}{\partial \rho'},\frac{\partial y}{\partial \rho'}\right)\right| = \sqrt{\cos^2\vartheta'+\lambda^{-2}\sin^2\vartheta'}, \\
		h_{\vartheta'} &= \left|\left(\frac{\partial x}{\partial \vartheta'},\frac{\partial y}{\partial \vartheta'}\right)\right| = \rho'\sqrt{\sin^2\vartheta' + \lambda^{-2}\cos^2\vartheta'},
	\end{align}
	therefore we can determine that for a scalar function $f$, and a vector $\vv=\left(v_1,v_2\right)$, we have
	\begin{equation}
		\nabla' f = \left(\frac{1}{h_{\rho'}}\frac{\partial f}{\partial \rho'}, \frac{1}{h_{\vartheta'}}\frac{\partial f}{\partial \vartheta'}\right)
	\end{equation}
	and
	\begin{equation}
		\nabla'\cdot\vv = \frac{1}{h_{\rho'} h_{\vartheta'}} \left[ \frac{\partial}{\partial \rho'} \left(v_1 h_{\vartheta'}\right) + \frac{\partial}{\partial \vartheta'}\left(v_2 h_{\rho'} \right)\right].
	\end{equation}
	From here we can determine the velocity field of the vortex as 
	\begin{equation}
		\vv = \nabla' S\left(\vartheta'\right) = \frac{1}{\rho'\sqrt{\sin^2\vartheta' + \lambda^{-2}\cos^2\vartheta'}}\frac{\partial S}{\partial \vartheta'} \boldsymbol{\hat{\vartheta}'},
	\end{equation}
	and therefore the stationary solutions to the continuity 
	
	\newpage 
	
	\begin{figure}[!t]
		\centering
		\includegraphics[width = \columnwidth]{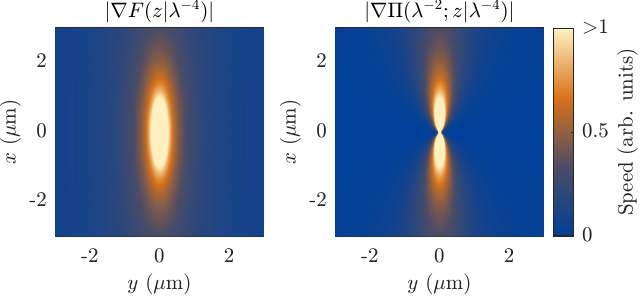}
		\caption{Contributions to the dipolar vortex velocity. Panels show the gradient of the corresponding terms appearing in Eq.~\eqref{eqn:analytic_phase} with $\lambda=2$.}
		\label{fig:speeds}
	\end{figure}

	\noindent equation are given by 
	\begin{equation}
		\frac{\partial}{\partial \vartheta'} \left[ \frac{n\left(\rho'\right)}{\rho'} \sqrt{\frac{\cos^2\vartheta' + \lambda^{-2}\sin^2\vartheta'}{\sin^2\vartheta' + \lambda^{-2}\cos^2\vartheta'}} \frac{\partial S}{\partial \vartheta'} \right] = 0.
	\end{equation}
	It is clear that $n(\rho')/\rho'$ can be removed, and so the remaining terms in the square brackets must be a constant. Rearranging this, and dropping the partial derivative, we have 
	\begin{equation}
		\frac{dS}{d\vartheta'} = c_1 \sqrt{\frac{\lambda^2\tan^2\vartheta'+1}{\lambda^2+\tan^2\vartheta'}},
	\end{equation}
	where $c_1$ is a constant. We integrate this to recover
	\begin{equation}
		S(x,y) = c_1 \left\{F\left( z\big| \lambda^{-4}\right) + (\lambda^{-2}-1)\Pi\left(\lambda^{-2}; z\big|\lambda^{-4}\right)\right\} +c_2,
		\label{eqn:analytic_phase}
	\end{equation}
	where $z(x,y) = i\text{arcsinh}(\lambda^2y/x)$ and $c_2$ is a constant. Here, $F(z|n)$ is the incomplete elliptic integral of the first kind, and $\Pi(m;z|n)$ is the incomplete elliptic integral of the third kind \cite{Abramowitz1988Handbook}. In this form, it is instructive to visualize each individual contribution, as shown in Fig.~\ref{fig:speeds}. The  incomplete elliptic integrals of the first and third kind have derivatives
	\begin{eqnarray}
		\frac{\partial F(z|n)}{\partial z} &=& \frac{1}{\sqrt{1-n\sin^2 z}}, \\
		\frac{\partial \Pi(m;z|n)}{\partial z} &=& \frac{1}{\left(1-m\sin^2 z\right)\sqrt{1-n\sin^2z}},
	\end{eqnarray}
	respectively. The constant $c_2$ is a phase shift, which can be set to zero. Transforming to Cartesian coordinates, we recover Eq.~\eqref{eqn:elliptic_phase} from the main text.
	When transforming from shear polar to Cartesian coordinates the second term in Eq.~\eqref{eqn:analytic_phase} changes sign.  The constant $\Lambda$ needs to be fixed such that $S(x,y)=\pm\pi$ when $\arctan(\lambda y/x)=\pm\pi$. This leads to 
	\begin{equation}
		\Lambda = \pi \left[ 4 K\left(1-\lambda^{-4}\right)-2\lambda^{-2}\Pi\left(1-\lambda^{-2}\big|1-\lambda^{-4}\right)\right]^{-1},
	\end{equation}
	where $K(n)=F(\pi/2,n)$ is the complete elliptic integral of the first kind. When evaluating Eq.~\eqref{eqn:elliptic_phase} numerically, one needs to be careful of the position of the branch-cut. In analogy to the $\texttt{atan2}$ function, we compute $S'(x,y)$, where \begin{align}
		S'(x,y) = \begin{cases}
			S(x,y) & \text{if } x > 0, \\
			S(x,y) - \pi & \text{if } x < 0 \text{ and } y \geq 0, \\
			S(x,y) + \pi & \text{if } x < 0 \text{ and } y < 0, \\
			-\frac{\pi}{2} & \text{if } x = 0 \text{ and } y > 0, \\
			+\frac{\pi}{2} & \text{if } x = 0 \text{ and } y < 0, \\
			\text{undefined} & \text{if } x = 0 \text{ and } y = 0.
		\end{cases}
	\end{align}
	
	We observe that in the limit where $\lambda=1$, the incomplete elliptic integrals $F\left[\arctan(y/x)\, | \,0\right]=\arctan(y/x)$ and $\Pi\left[0\, ; \arctan\left(y/x\right)\, |\, 0\right]=\arctan(y/x)$, while the complete elliptic integrals reduce to $  K(0)=\Pi(0,0)=\pi/2$. This means that Eq.~\eqref{eqn:elliptic_phase} reduces to the non-dipolar case, $S(x,y)=q\arctan(y/x)$ \cite{Fetter1966a}. 
	
	\begin{figure}
		\centering
		\includegraphics[width = \columnwidth]{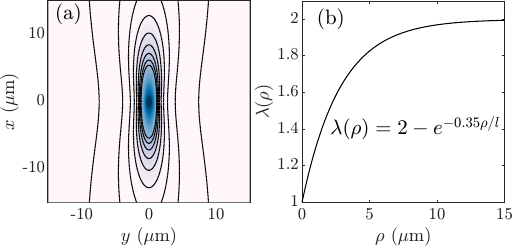}
		\caption{Analytic phase with spatially varying $\lambda$. (a) Speed from Eq.~\eqref{eqn:elliptic_phase} with $\lambda(\rho)$, with same parameters as Fig.~\ref{fig:one_vortex}(f). (b) Function used in (a), with the length scale $l=1$\textmu{m}.}
		\label{fig:lambdarho}
	\end{figure}

	{\it Appendix B}---Here, we discuss the optimal value of $\lambda$, and show that to obtain the best agreement with the GPE velocity field, one should take $\lambda(\rho)$. The necessity for this can be seen by comparing Figs.~\ref{fig:one_vortex}(e) and (f), where the constant value of $\lambda$ that provides excellent agreement far from the vortex core overestimates the ellipticity close to the core. In Fig.~\ref{fig:lambdarho}, we show that both the core and far field can be captured by taking a radially dependent $\lambda$, shown exemplarily for the case $(\edd,\alpha) = (0.9,\pi/2)$. Note that the final value $\lambda \to 2$ is sensitive to the GPE parameters and is modified if we vary our units. Such an observation may be useful for future applications of our phase solution when trying to closely match results from the GPE.
	
	We note here that in Figs.~\ref{fig:two_vortices} and \ref{fig:three_vortices}, the value of $\lambda$ appears lower than the optimized values taken for a single vortex (Figs.~\ref{fig:one_vortex} and \ref{fig:lambdarho}). We attribute this to vortex–vortex interactions that modify the shape of the vortex core, and note that implementations of the DPVM typically require only $\lambda \in [1, 1.3]$ to capture the physics of the GPE.
	
	{\it Appendix C}---Our approach relies on a dimensional reduction of the full three-dimensional wavefunction, which we express as $\Psi(\mathbf{r},t)\equiv\Psi(\boldsymbol{\rho},z,t)=\psi(\boldsymbol{\rho},t)\chi_\sigma(z)$, where $\boldsymbol{\rho}=(x,y)$ and $\chi_\sigma(z) = \pi^{-1/4}\sigma^{-1/2}e^{-z^2/2\sigma^2}$ represents a Gaussian profile along the $z$-axis with a variational width $\sigma$. By substituting this Ansatz into the three-dimensional Gross-Pitaevskii energy functional and integrating over the axial coordinate $z$, we derive an effective energy functional of the form $E=E_\perp+E_z$, where
	\begin{align}
		E_\perp = \int\text{d}^2\boldsymbol{\rho}\,\psi^*\left[-\frac{\hbar^2\nabla^2}{2m} +\frac{1}{2}\Phi\right]\psi\,,
		\label{eqn:energy_perp}
	\end{align}
	and the axial contribution is given by $E_z=\hbar\omega_z(l_z^2/4\sigma^2 + \sigma^2/4l_z^2)$, with $l_z = \sqrt{\hbar/(m\omega_z)}$ denoting the harmonic oscillator length along $z$. This equation has been used to calculate the vortex energy in Fig.~\ref{fig:one_vortex}(d).
	
	The effective two-dimensional interaction potential $\Phi(\boldsymbol{\rho},t)$ is conveniently computed in Fourier space:
	\begin{align}
		\Phi(\boldsymbol{\rho},t) = \mathcal{F}^{-1}\left[\tilde{U}_{\alpha,\sigma}(\boldsymbol{k})\mathcal{F}\left[|\psi(\boldsymbol{\rho},t)|^2\right]\right]\,,
	\end{align}
	where $\boldsymbol{k}=(k_x,k_y)$, and $\mathcal{F}$ ($\mathcal{F}^{-1}$) denotes the (inverse) Fourier transform. The Fourier-transformed interaction kernel takes the form
	\begin{align}
		\tilde{U}_{\alpha,\sigma}(\boldsymbol{q}) = \frac{\sqrt{2\pi}\hbar^2}{\sigma m}\left(a_s+a_\text{dd}\left[F_\parallel(\boldsymbol{q})\sin^2\alpha+F_\perp(\boldsymbol{q})\cos^2\alpha\right]\right)\,,
	\end{align}
	with $\boldsymbol{q}=\boldsymbol{k}\sigma/\sqrt{2}$, and the angular-dependent functions defined as $F_\parallel(\boldsymbol{q}) = -1 + 3\sqrt{\pi}\frac{q_x^2}{q}e^{q^2}\text{erfc}(q)$ and $F_\perp(\boldsymbol{q}) = 2 - 3\sqrt{\pi}qe^{q^2}\text{erfc}(q)$. This expression corresponds to the dimensionally reduced version of $U(\textbf{r})$.
	
	By taking the functional derivative $\delta E_\perp/\delta\psi^*$, one recovers the effective two-dimensional Gross-Pitaevskii equation presented in Eq.~\eqref{eqn:2DGPE} of the main text.
	
	We solve the Gross-Pitaevskii equation for an infinite planar system using a fourth-order Runge-Kutta method with Dirichlet boundary conditions fixed at uniform density. Simulations are performed on a $512\times512$ grid over a $50\times50\,\mu\text{m}^2$ domain. For the extreme case $\alpha=\pi/2$, results were consistent on a $1024\times1024$ grid in a $100\times100\,\mu\text{m}^2$ box. Initial conditions are prepared by phase imprinting vortices with azimuthal symmetry, followed by imaginary time evolution. A short evolution in complex time further reduces noise, and absorbing boundaries suppress edge reflections.

\end{document}